\documentclass[a4paper]{jpconf}
\usepackage{graphicx}
\usepackage{amsmath}

\newcommand{\newc}{\newcommand}
\newc{\be}{\begin{equation}}
\newc{\ee}{\end{equation}}
\newc{\bea}{\begin{eqnarray}}
\newc{\eea}{\end{eqnarray}}

\begin{document}
\title{CP-odd invariants for multi-Higgs models and
applications with discrete symmetry}

\author{Ivo de Medeiros Varzielas}

\address{CFTP, Departamento de F\'{\i}sica, Instituto Superior T\'{e}cnico,\\
Universidade de Lisboa,
Avenida Rovisco Pais 1, 1049 Lisboa, Portugal}

\ead{ivo.de@udo.edu}

\begin{abstract}

CP-odd invariants are useful for studying the CP properties of Lagrangians in any basis.
We explain how to build basis invariants for the scalar sector, and how to distinguish CP-odd invariants from CP-even invariants. Up to a certain order, we use these methods to systematically build all the CP-odd invariants. 
The CP-odd invariants signal either explicit or spontaneous violation of CP. Making use of the CP-odd invariants, we determine the CP properties of
potentials with 3 and with 6 Higgs fields arranged as triplets of specific discrete symmetries in the $\Delta(3n^2)$ or $\Delta(6n^2)$ series (inlcuding $A_4$, $S_4$, $\Delta(27)$ and $\Delta(54)$ as well as the cases for $n>3$).

\end{abstract}

\section{Introduction}

This contribution to the proceedings of DISCRETE 2016 is based on \cite{Varzielas:2016zjc}.

CP violation (CPV) is present in the Standard Model (SM) in the Yukawa sector, due to the replication of fermion in 3 generations. Physics does not depend on choice of basis, and indeed the amount of CPV in the SM can be calculated in a weak basis invariant which can be expressed in terms of the quark mass matrices \cite{Jarlskog:1985ht, Bernabeu:1986fc}.
In the SM scalar sector, with a single Higgs doublet, the possibility of CPV does not exist. CPV is enabled in the scalar sector of the 2 (or more) Higgs Doublet Model (HDM) \cite{Lee:1973iz}. The extension of the scalar content of the SM is well motivated, as additional CPV sources are required to account for the baryon asymmetry of the Universe \cite{Sakharov:1967dj,Kuzmin:1985mm}. In the scalar sector, CPV can appear explicitly or through the vacuum expectation values (VEVs), the latter case consisting in Spontaneous CPV (SCPV).

There is often some subtletly to the CP properties of theories (e.g. it is possible to have complex parameters in a Lagrangian and for it to conserve CP), which increases the appeal of an invariant approach, and the generalisation of the powerful invariant approach to CP~\cite{Bernabeu:1986fc} to theories with discrete symmetries recently lead to other relevant CP-odd invariants (CPIs) for the respective Yukawa sector ~\cite{Branco:2015hea, Branco:2015gna}. The invariant approach to CP as applied to scalar potentials in~\cite{Mendez:1991gp,Lavoura:1994fv, Botella:1994cs, Branco:2005em, Davidson:2005cw, Gunion:2005ja} was further developed in \cite{Varzielas:2016zjc}.

In order to find CPIs for arbitrary potentials, we use either the kind of diagrams introduced for the 2HDM in~\cite{Davidson:2005cw} or a new method based on ``contraction matrices'', and catalogued several new CPIs. We then apply them to several potentials with 3 or 6 Higgs fields with a discrete symmetry from the $\Delta(3n^2)$ and $\Delta(6n^2)$ series. These symmetries are considered frequently in the context of flavour and CP models \cite{
Feruglio:2012cw,
Ding:2013hpa, Feruglio:2013hia, King:2013vna, Luhn:2013vna, Ding:2013bpa, Li:2013jya, Ding:2013nsa,
King:2014rwa, Ding:2014ssa, Ding:2014hva, Neder:2014mxa, Li:2014eia, Hagedorn:2014wha, Ding:2014ora,
Bjorkeroth:2015ora, Li:2015jxa, DiIura:2015kfa, Ballett:2015wia, Neder:2015wsa, Turner:2015uta, Ding:2015rwa, Bjorkeroth:2015uou,
Li:2016ppt}.

In particular, the 3HDM \cite{Branco:1983tn, Toorop:2010ex, Toorop:2010kt, deMedeirosVarzielas:2011zw, Varzielas:2012nn, Bhattacharyya:2012pi, Ivanov:2012fp,Ivanov:2012ry, Varzielas:2013sla, Varzielas:2013eta, Keus:2013hya, Keus:2014jha, Ivanov:2014doa, Keus:2015xya, Fallbacher:2015rea, Ivanov:2015mwl, Emmanuel-Costa:2016vej} and the 6HDM \cite{Ivanov:2013nla,Ivanov:2010ww,Ivanov:2010wz,Hernandez:2013dta, Keus:2014isa, Nishi:2014zla, Varzielas:2015joa} where the 3 or 6 Higgses are related by a discrete symmetry as one or two (flavour) triplets have been extensively studied.
Nevertheless, our systematic study through the new CPIs revealed several novel results.

\section{CP-odd invariants for potentials}

In our notation, we write scalar potentials in a standard form. This is similar to the notation in~\cite{Branco:2005em,Davidson:2005cw} but with some important differences. An even potential of $N$ scalar fields $\varphi_i$, where $\phi=(\varphi_1,\ldots , \varphi_N)$ and $\phi^\ast=(\varphi_1^\ast,\ldots , \varphi_N^\ast)$, can be rewritten as
\begin{equation}
V~=~ {\phi^\ast}^a  \,Y_a^b\, \phi_b 
+  {\phi^\ast}^a{\phi^\ast}^c  \,Z_{ac}^{bd}\, \phi_b \phi_d \ .
\label{eq:potCPgeneral} 
\end{equation}
The lower indices on $Y$ and $Z$ are
contracted with $\phi^\ast$ and upper indices with $\phi$. 
The tensors $Y$ and $Z$ contain the couplings from the potential (the formalism can be extended to account also for tri-linear terms). $Z^{bd}_{ac}$ is by definition unchanged if $a\leftrightarrow c$ or
$b\leftrightarrow d$.

In particular, potentials of Higgs doublets are rewritten in the standard form with their components: $n$ Higgs doublets $H_{i
  \alpha}=(h_{i,1},h_{i,2})$, with $\alpha=1,2$ for $SU(2)_L$ index and $i$ identifies the doublet from $1$ to $n$, are recast as
\begin{equation}
\phi=(\varphi_1,\varphi_2, \ldots, \varphi_{2n-1},
\varphi_{2n})=(h_{1,1},h_{1,2},\ldots,h_{n,1},h_{n,2}) \ ,
\label{eq:phi_to_h}
\end{equation}

The invariance of the potential under whatever symmetries it is invariant under (e.g. $SU(2)_L\times U(1)_Y$) will be manifest in the structure of the tensors $Y$ and $Z$:
if under $G$, $\phi$ transforms in some representation $\rho(g)$, $g\in G$,
\begin{eqnarray}
 \phi_a&\mapsto& [\rho(g)]_{a}^{a'} \phi_{a'}\ ,\\[2mm]
 \phi^{\ast a}&\mapsto &\phi^{\ast a'} [\rho^\dagger(g)]^{a}_{a'}\ ,
\end{eqnarray}
then
\begin{equation}
Y_a^b =  \rho_a^{a'} \, Y_{a'}^{b'} \, {\rho^\dagger}_{b'}^b \ ,
\end{equation}
\begin{equation}
Z_{ac}^{bd} =  \rho_a^{a'}\, \rho_c^{c'} \, Z_{a'c'}^{b'd'} \, 
{\rho^\dagger}_{b'}^b \,{\rho^\dagger}_{d'}^d \ ,
\label{symm_of_Z}
\end{equation}
where we denote $ \rho_a^{a'}= [\rho (g)]_a^{a'}$.

An arbitrary basis transformation acts on the fields by a unitary
$N\times N$ matrix
\bea
\phi_a & \mapsto & V_a^{a'} \phi_{a'}  \ , \\
{\phi^\ast}^a &\mapsto &  {\phi^\ast}^{a'} {V^\dagger}_{a'}^a \ ,
\eea
which changes tensor components accordingly:
\bea
Y_a^b &\mapsto & V_a^{a'} \, Y_{a'}^{b'} \, {V^\dagger}_{b'}^b \ ,\\
Z_{ac}^{bd} &\mapsto & V_a^{a'}\, V_c^{c'} \, Z_{a'c'}^{b'd'} \, 
{V^\dagger}_{b'}^b \,{V^\dagger}_{d'}^d \ .
\eea

In our notation, complex conjugation changes indices of a field:
\bea
\phi_a & \mapsto & (\phi_a)^\ast \equiv  {\phi^\ast}^a  \ ,\\
{\phi^\ast}^a  & \mapsto &  ({\phi^\ast}^a)^\ast \equiv {\phi}_a \ ,
\eea
therefore,
\bea
{\phi^\ast}^a  \,Y_a^b \, \phi_b ~\mapsto~
 {\phi}_a  \,(Y_a^b)^\ast\, {\phi^\ast}^b =
{\phi^\ast}^b  \,(Y_a^b)^\ast\, {\phi}_a  ={\phi^\ast}^a  \,(Y_b^a)^\ast\, {\phi}_b \ ,
\eea
and we obtain (because $V^\ast =V$):
\be
(Y_b^a)^\ast ~=~ Y_a^b \ .
\label{Y_real}
\ee 
Similarly, for the $Z$ tensor we have
\be
(Z^{ac}_{bd})^\ast ~=~ Z_{ac}^{bd} \ .
\label{Z_real}
\ee

A general CP transformation acts with a unitary matrix $U$:
\bea
\phi_a &\mapsto & {\phi^\ast}^{a'} U_{a'}^a \ , \\
{\phi^\ast}^a &\mapsto & {U^\dagger}_a^{a'} \phi_{a'} \ ,
\eea
and thus
\bea
{\phi^\ast}^a  \,Y_a^b \, \phi_b ~\mapsto~
{U^\dagger}_a^{a'} \phi_{a'}   \,Y_a^b\,  
{\phi^\ast}^{b'} U_{b'}^b   &=&
{U^\dagger}_a^{a'} \phi_{a'}   \,(Y_b^a)^\ast\,  
{\phi^\ast}^{b'} U_{b'}^b  \\
&=& 
{\phi^\ast}^{b'} U_{b'}^b 
\,(Y_b^a)^\ast\,  
{U^\dagger}_a^{a'} \phi_{a'} \\
&=& 
{\phi^\ast}^{a} U_{a}^{a'} 
\,(Y_{a'}^{b'})^\ast\,  
{U^\dagger}_{b'}^{b} \phi_{b} \ .
\eea
A comparison of this with the original term, and the respective exercise for $Z$, reveals how the general CP transformation acts on the tensors:
\bea
Y_a^b &\mapsto & U_{a}^{a'} 
\,(Y_{a'}^{b'})^\ast\,  
{U^\dagger}_{b'}^{b} \ , \label{eq:CPcond1}\\
Z_{ac}^{bd} &\mapsto & U_{a}^{a'}  U_{c}^{c'} 
\,(Z_{a'c'}^{b'd'})^\ast\,  
{U^\dagger}_{b'}^{b} {U^\dagger}_{d'}^{d}\ .\label{eq:CPcond2}
\eea
This allows us to recast the condition for the potential to be CP invariant: if there is any $U$ such that the left- and right-hand sides  of
  Eqs.~(\ref{eq:CPcond1}) and (\ref{eq:CPcond2}) are identical.  
The trivial CP:
\be
U^{a}_{a'} =  \delta^{a}_{a'}\ .
\label{CP0}
\ee
obeys the condition for real $Y$ and $Z$ tensors. We refer to this trivial CP throughout as $CP_{0}$.

When considering SCPV, we build spontaneous CPIs (SCPIs) by using also VEVs to form basis invariants. It is therefore useful to note they transform under basis change as:
\bea
 v_a&\mapsto &V_a^{a'}v_{a'} \ ,\label{eq:basi}\\
 {v^\ast}^a&\mapsto &v^{\ast a'}\,{V^\dagger}^{a}_{a'}\ ,\label{eq:basi*}
\eea
where $v\equiv(v_1,\ldots)$, $v_i=\langle\varphi_i\rangle$. Under general CP transformations:
\bea
 v_a&\mapsto& v^{\ast a'}U^{a}_{ a'}\ , \\
 {v^\ast}^a&\mapsto&   {U^\dagger}_a^{a'}v_{a'} \ .
\eea

\subsection{Simplest basis invariants}

If one has a combination of $Y$ and $Z$ tensors with all indices contracted, it forms a basis invariant. Thus, the simplest invariant is the Trace of $Y$:
\begin{equation}
 Y^a_a.
 \label{1Y_invariant}
\end{equation}
The possibilities with two $Y$ tensors are:
\begin{equation}
Y^a_a Y^b_b\,, \quad Y^a_bY^b_a.
\end{equation}
We note that these contractions can be mapped to elements of the permutation group by identifying the permutation that takes the ordering in the upper indices to the ordering in the lower indicies:
\begin{equation}
Y^a_a Y^b_b \Leftrightarrow a\mapsto a \,, \quad b\mapsto b \ ,
\end{equation}
and
\begin{equation}
Y^a_b Y^b_a \Leftrightarrow a\mapsto b \,, \quad b\mapsto a\ .
\end{equation}
We write thus:
\begin{equation}
 Y^a_{\sigma(a)}Y^b_{\sigma(b)} \,, \quad \sigma \in S_2 \ ,
\end{equation}
where $\sigma$ is an element of the group $S_2$. Out of the two basis invariants with two $Y$ tensors, one is simply the square of the Trace of $Y$, i.e. it can be expressed in terms of smaller basis invariants. We refer to basis invariants that can't be written in terms of smaller ones as irreducible.

Similarly for $Z$ tensors, the simplest invariants are:
\begin{equation}
 Z^{ab}_{\sigma(a)\sigma(b)} \,, \quad \sigma \in S_2\ ,
\end{equation}
or explicitly:
\begin{equation}
Z^{ab}_{ab}\,, \quad Z^{ab}_{ba}\ .
\end{equation}
Due to the inherent symmetry $Z$ tensors have under exchange of upper (or lower) indices, these two basis invariants are the same.
With two $Z$ tensors, the possible contractions are 24:
\begin{equation}
Z^{ab}_{\sigma(a)\sigma(b)}Z^{cd}_{\sigma(c)\sigma(d)} \,, \quad \sigma
\in S_4\ ,
\end{equation}
but there are only two new irreducible basis invariants, which can be taken to be
\begin{equation}
Z^{ab}_{bd}Z^{cd}_{ac} \,, \quad Z^{ab}_{cd}Z^{cd}_{ab}.
\label{2Z_invariants}
\end{equation}

In general, a basis invariant $I_\sigma^{(n_Z,m_Y)}$ with $m_Y$ Y tensors and $n_Z$ $Z$ tensors is
\begin{equation}
I_\sigma^{(n_Z,m_Y)}\equiv Y^{a_1}_{\sigma(a_1)}\ldots Y^{a_{m_Y}}_{\sigma(a_{m_Y})}Z^{b_1 b_2}_{\sigma(b_1)\sigma(b_2)}\ldots Z^{b_{2n_Z-1} b_{2n_Z}}_{\sigma(b_{2n_Z-1})\sigma(b_{2n_Z})} \,, \quad \sigma \in S_{m_Y+2n_Z}.
\label{invariant_definition}
\end{equation}

Many basis invariants are not CP-odd. Under a general CP transformation, a coupling
tensor goes to its complex conjugate and is acted on by unitary transformations $U$. Being basis invariant means the $U$ matrices drop out, so within a basis invariant, a general CP transformation reduces to converting tensors into their complex conjugates (by exchanging upper and lower indices). For example, the Trace of $Y$:
\begin{equation}
 Y^a_a \rightarrow{CP} (Y^{a'}_{a''})^\ast {U^\dagger}_{a'}^aU^{a''}_a=(Y^{a'}_{a''})^\ast \delta_{a'}^{a''}=(Y^a_a)^\ast=Y^a_a.
\end{equation}
Likewise:
\bea
I_\sigma^{(n_Z,m_Y)} \equiv Y^{a_1}_{\sigma(a_1)}\ldots Y^{a_{m_Y}}_{\sigma(a_{m_Y})}Z^{b_1 b_2}_{\sigma(b_1)\sigma(b_2)}\ldots Z^{b_{2n_Z-1} b_{2n_Z}}_{\sigma(b_{2n_Z-1})\sigma(b_{2n_Z})} \nonumber\\ \rightarrow{CP}
Y^{\sigma(a_1)}_{a_1}\ldots
Y^{\sigma(a_{m_Y})}_{a_{m_Y}}Z^{\sigma(b_1)\sigma(b_2)}_{b_1 b_2}\ldots
Z^{\sigma(b_{2n_Z-1})\sigma(b_{2n_Z})}_{b_{2n_Z-1} b_{2n_Z}}=
[I_\sigma^{(n_Z,m_Y)}]^{\ast}\ .
\label{CP_of_invariant}
\eea
From a basis invariant $I$ differing from its CP conjugate $I^*$, we can build a CPI $ \mathcal{I}$:
\begin{equation}
 \mathcal{I}=I-I^{\ast}.
 \label{eq:CPI_I_c}
\end{equation}

In order to better find useful CPIs we employ diagrams and in particular contraction matrices that reveal which basis invariants can form CPIs.

For the diagrams, the rules are as follows
 \begin{equation}
X^{a.}_{..}X^{..}_{a.}=\vcenter{\hbox{\includegraphics[scale=0.2]{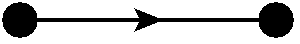}}}
\end{equation}
Lines don't need to be distinguished:
\begin{equation}
Z^{ab}_{..}Z^{..}_{ab}=\includegraphics[scale=0.2]{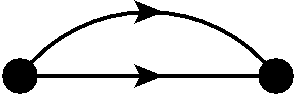}
\label{FR_2_definition}
\end{equation}
Contracting indices on same tensor:
\begin{equation}
X^{a.}_{a.}=\vcenter{\hbox{\includegraphics[scale=0.2]{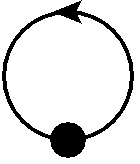}}}
\label{loop_diagram}
\end{equation}
For the contraction matrices, we list all the $Y$ tensors, then all the $Z$ tensors as rows and columns and have $0,1,2...$ if there are $0,1,2,...$ lines connecting from upper indices of that row's tensor to the lower index of that column's tensor. Some examples are:

\begin{equation}
 Y^a_a=\includegraphics[scale=0.2]{diagrams/FR_3.png}=\begin{pmatrix}1\end{pmatrix}
\end{equation}

\begin{equation}
 Y^a_aY^b_b=\includegraphics[scale=0.2]{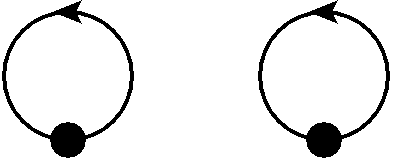}=\begin{pmatrix}1&0\\0&1\end{pmatrix}
\end{equation}
\begin{equation}
 Y^a_b Y^b_a=\includegraphics[scale=0.2]{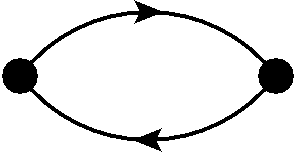}=\begin{pmatrix}0&1\\1&0\end{pmatrix}
\end{equation}

\begin{equation}
 Z^{ab}_{ab}=\includegraphics[scale=0.2]{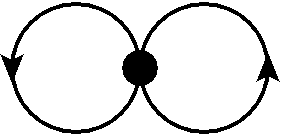}=\begin{pmatrix}2\end{pmatrix}
\end{equation}
\begin{equation}
 Z^{ac}_{bc}Y^b_a=\includegraphics[scale=0.2]{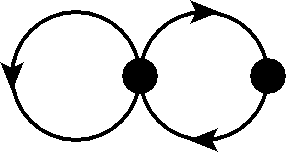}=\begin{pmatrix}0&1\\1&1\end{pmatrix}
\end{equation}
\begin{equation}
 Z^{ab}_{cd}Z^{cd}_{ab}=\includegraphics[scale=0.2]{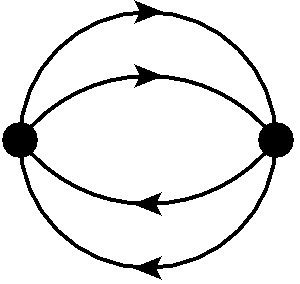}=\begin{pmatrix}0&2\\2&0\end{pmatrix}
\end{equation}
\begin{equation}
 Z^{ab}_{ac}Z^{cd}_{bd}=\includegraphics[scale=0.2]{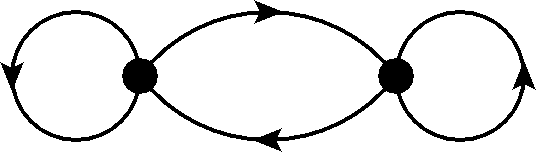}=\begin{pmatrix}1&1\\1&1\end{pmatrix}
\end{equation}

A CPI can be built from a basis invariant that changes if all upper indices are exchanged with the lower indices, which, up to rearranging the tensors, corresponds to diagrams that are distinct when the arrows are reversed and to contraction matrices that not symmetric (i.e. are distinct by exchanging rows and columns). The smallest CPI can be built from the following basis invariant:
\begin{equation}
 I_1\equiv Z^{ab}_{ae}Z^{cd}_{bf}Y^e_cY^f_d=\includegraphics[scale=0.2]{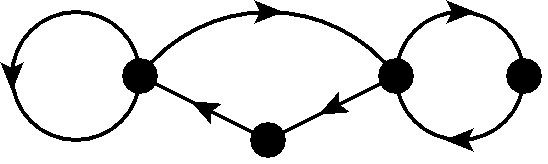}=\begin{pmatrix}0&0&1&0\\0&0&0&1\\0&0&1&1\\1&1&0&0\end{pmatrix}
\end{equation}
and its distinct CP conjugate
\begin{equation}
I_1^{*}\equiv Z^{ae}_{ab}Z^{bf}_{cd}Y^c_eY^d_f=\vcenter{\hbox{\includegraphics[scale=0.2]{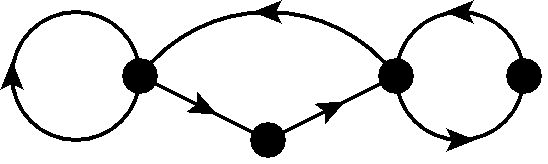}}}
\label{I21CP}
\end{equation}
where in the matrix the $Y$ tensors are the first rows and columns (even though they are ordered differently in the expressions and diagrams).

For the purpose of the potentials considered, the useful CPIs are order 6 in $Z$ tensors:
\begin{equation}
I_2^{(6)}=Z^{a_1 a_2}_{a_7 a_{10}} Z^{a_3 a_4}_{a_{11} a_6} Z^{a_5 a_6}_{a_9 a_8} Z^{a_7 a_8}_{a_3 a_{12}} Z^{a_9 a_{10}}_{a_5 a_4} Z^{a_{11} a_{12}}_{a_1 a_2}=\vcenter{\hbox{\includegraphics[scale=0.2]{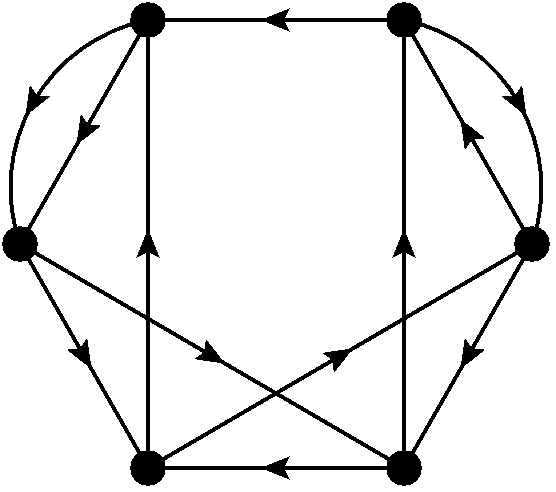}}}=\left(
\begin{array}{cccccc}
 0 & 0 & 0 & 0 & 0 & 2 \\
 0 & 0 & 0 & 1 & 1 & 0 \\
 0 & 1 & 0 & 0 & 1 & 0 \\
 1 & 0 & 1 & 0 & 0 & 0 \\
 1 & 0 & 1 & 0 & 0 & 0 \\
 0 & 1 & 0 & 1 & 0 & 0 \\
\end{array}
\right)
\end{equation}
\begin{equation}
 I_4^{(6)}=Z^{a_1 a_2}_{a_{11} a_{10}} Z^{a_3 a_4}_{a_5 a_8} Z^{a_5 a_6}_{a_7 a_{12}} Z^{a_7 a_8}_{a_9 a_6} Z^{a_9 a_{10}}_{a_1 a_4} Z^{a_{11} a_{12}}_{a_3 a_2}=\vcenter{\hbox{\includegraphics[scale=0.2]{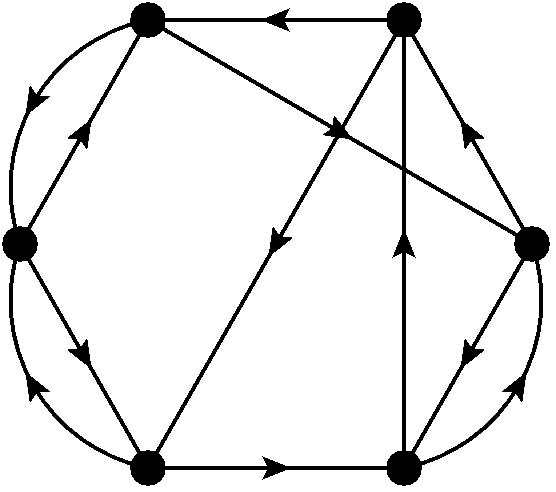}}}=\left(
\begin{array}{cccccc}
 0 & 0 & 0 & 0 & 1 & 1 \\
 0 & 0 & 0 & 0 & 1 & 1 \\
 0 & 1 & 0 & 1 & 0 & 0 \\
 0 & 1 & 1 & 0 & 0 & 0 \\
 1 & 0 & 0 & 1 & 0 & 0 \\
 1 & 0 & 1 & 0 & 0 & 0 \\
\end{array}
\right)
\end{equation}
A full list of CPIs up to order 6 in $Z$ tensors can be found in the appendix of \cite{Varzielas:2016zjc}.

In addtion, SCPIs can be built by contracting in a similar fashion VEVs (which transform under basis changes and general CP transformations as the respective field does). The SCPI that is relevant for the potentials considered is:
\begin{equation}
J_{1}^{(3,2)}\equiv Z^{a_1a_2}_{a_4a_5}Z^{a_3a_4}_{a_2a_6}Z^{a_5a_6}_{a_7a_8}v_{a_1}v_{a_3}v^{\ast a_7}v^{\ast a_8}=\vcenter{\hbox{\includegraphics[scale=0.2]{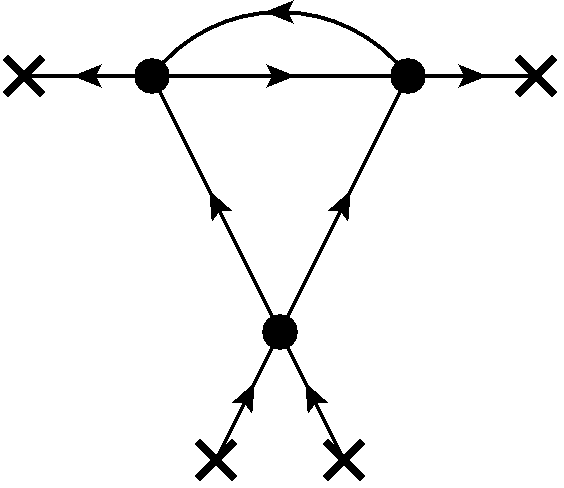}}}
\label{J322}
\end{equation}

\section{Explicit CP violation}

The potential of one triplet of $\Delta(27)$ has interesting CP properties. We write
\bea
V_0 (\varphi) =
 - ~m^2_{\varphi}\sum_i   \varphi_i \varphi^{*i}
+ r \left( \sum_i   \varphi_i \varphi^{*i}  \right)^2
+ s \sum_i ( \varphi_i \varphi^{*i})^2
\eea

\bea
V_{\Delta(27)} (\varphi)=
V_0(\varphi)
+ \left[d \left(
\varphi_1 \varphi_1 \varphi^{*2} \varphi^{*3} + 
\text{cycl.} \right) +\text{h.c.}\right]
\eea
for $SU(2)_L$ singlet scalars, where we consider an additional $U(1)$ flavour symmetry charging the triplet to exclude tri-linear terms, making the potential similar to the respective potentials for $SU(2)_L$ doublets
\bea
 V_{0} ( H ) &=&
 - ~m^2_{h}\sum_{i, \alpha}   h_{i \alpha}  h^{*i\alpha}
+\sum_{i, j, \alpha, \beta} \left[ r_1 ( h_{i \alpha}  h^{*i\alpha})( h_{j \beta}  h^{*j \beta}) + r_2 ( h_{i \alpha}  h^{*i \beta})( h_{j \beta}  h^{*j\alpha}) \right]\notag \\
&&+ s \sum_{i, \alpha, \beta}  ( h_{i \alpha}  h^{*i\alpha})( h_{i \beta}
h^{*i\beta})
\eea

\bea
V_{\Delta(27)} (H) = V_0(H) 
~+~ \sum_{\alpha, \beta} \left[d \left(
h_{1 \alpha} h_{1 \beta}  h^{*2 \alpha}  h^{*3 \beta} + 
\text{cycl.} \right) +
\text{h.c.}\right]
\eea

We consider also the potential with two triplets of $\Delta(3n^2)$ for $n>3$. We write:
\bea
V_1 (\varphi,\varphi') &=&
+~ \tilde r_1 \left( \sum_i \varphi_i \varphi^{*i} \right)
\left( \sum_j \varphi'_j \varphi'^{*j} \right) 
+ \tilde r_2\left( \sum_i \varphi_i \varphi'^{*i} \right)
\left( \sum_j \varphi'_j \varphi^{*j} \right) \notag \\[2mm]
&& +~ \tilde s_1\sum_i \left(\varphi_i \varphi^{*i} \varphi'_i \varphi'^{*i}
\right) \notag \\[2mm]
&& +~ \tilde s_2 \left(
\varphi_1 \varphi^{*1} \varphi'_2 \varphi'^{*2} + 
\varphi_2 \varphi^{*2} \varphi'_3 \varphi'^{*3} + 
\varphi_3 \varphi^{*3} \varphi'_1 \varphi'^{*1} 
\right)  \notag \\[2mm]
&& +~ i \, \tilde s_3 
\Big[
(\varphi_1 \varphi'^{*1} \varphi'_2 \varphi^{*2} + \text{cycl.}
) 
- 
( \varphi^{*1}\varphi'_1  \varphi'^{*2} \varphi_2 +\text{cycl.}
)
\Big] .
\label{eq:potV1}
\eea

\bea
V_{\Delta(3n^2)} (\varphi,\varphi') &=&
V_0 (\varphi) + V_0'(\varphi') + V_1 (\varphi, \varphi') \ .
\label{eq:potDelta3n2-2b}
\eea

Explicit CPV is confirmed for these potentials as at least one CPI is non-zero. For $SU(2)_L$ singlets and $\Delta(27)$:
\begin{equation}
\mathcal{I}^{(6)}_{4} =  -\frac{3}{32} \left(d^3-d^{*3} \right) \left(d^3+6 d d^* s+ d^{*3}-8 s^3\right)
\label{D27CPI}
\end{equation}
and for $SU(2)_L$ doublets the expression is the same apart from a different fraction. The 6 solutions of $\mathcal{I}^{(6)}_{4} = 0$ in terms of $d$ and $s$ (parameters of the $\Delta(27)$ potential) correspond to imposing one of the CP symmetries identified in \cite{Nishi:2013jqa}, which naturally eliminates the possibility of explicit CPV.

For $\Delta(3n^2)$ we calculate:
\begin{equation}
\mathcal{I}^{(6)}_2 =  \frac{3}{512} i \tilde{s}_2 \tilde{s}_3 (-3 \tilde{r}_2^2 + \tilde{s}_3^2) [-\tilde{s}_1^2 + \tilde{s}_1 \tilde{s}_2 + 
   \tilde{r}_2 (-2 \tilde{s}_1 + \tilde{s}_2) + \tilde{s}_3^2]
\label{eq:inv3n2-2flavons}
\end{equation}
The solutions for $\mathcal{I}^{(6)}_{2}=0$ match CP symmetries that can be imposed on the potential:
e.g. trivial $CP_0$ forces $\tilde{s}_3=0$; another CP symmetry forces $\tilde{s}_2=0$ and after finding
$(-3 \tilde{r}_2^2 + \tilde{s}_3^2)$ in other CPIs, we suspected there was a CP symmetry CP forcing that combination of parameters to vanish and indeed found it.

A summary table with our results for explicit CPV can be found in \cite{Varzielas:2016zjc}.

\section{Spontaneous CP violation}

The $\Delta(27)$ invariant potentials has interesting SCPV properties, as its SCPV can be geometric~\cite{Branco:1983tn, deMedeirosVarzielas:2011zw, Varzielas:2012nn, Bhattacharyya:2012pi, Varzielas:2013sla, Varzielas:2013eta, Fallbacher:2015rea}. We analysed it with a SCPI and obtained
\bea
  && \mathcal{J}_1^{(3,2)} = \frac{1}{4}(d^{\ast 3}-d^3) Q(|v_i|)
  \\&&+\frac{1}{2}(dd^{\ast 2}-2d^\ast s^2+d^{2}s)(v_2 v_3 v_1^{\ast 2}+v_1
  v_3 v_2^{\ast2}+v_1 v_2 v_3^{\ast2}) \nonumber\\
&&-\frac{1}{2}(d^2d^\ast-2ds^2+d^{\ast 2}s)(v_2^\ast v_3^\ast
 v_1^2+v_1^\ast v_3^\ast v_2^2+v_1^\ast v_2^\ast v_3^2)   \ .\label{eq:J27ex2}
\eea
As the potential has explicit CPV, SCPV only applies after imposing a CP symmetry, which simplifies the expression. E.g., for $CP_0$ (forces $\text{Arg}(d)=0$):
\bea
&\mathcal{J}_1^{(3,2)} = \frac{1}{2}(d^3 -2d s^2+d^{2}s) \\
&\left[(v_2 v_3 v_1^{\ast 2}+v_1 v_3 v_2^{\ast2}+v_1 v_2 v_3^{\ast2})-(v_2^\ast v_3^\ast v_1^2+v_1^\ast v_3^\ast v_2^2+v_1^\ast v_2^\ast v_3^2) \right]
\eea
The results obtained through the SCPI matched the known results. For example, for $CP_0$, it can be checked from the dependence on the VEVs that $\langle \varphi \rangle = (1,\omega, \omega^2)$ makes the SCPI vanish (and indeed this VEV preserves a CP symmetry present in the potential when $CP_0$ is imposed). Conversely, it confirms the VEV $\langle \varphi \rangle = (\omega,1,1)$ has SCPV, as the SCPI is non-vanishing.

\section{Conclusion}

In summary, we developed the formalism for finding CP-odd basis invariants for scalar potentials, and performed a systematic search finding several new such invariants (up to 6 $Z$ tensors).
These methods for studying explicit and spontaneous CP violation are valid for any potential when brought to standard form, for any $SU(2)_L$ assignments of the scalars, and are therefore very useful in studying the CP properties of potentials in multi-Higgs models.
We exemplified the use of the CP-odd invariants by considering 3 and 6 Higgs doublet models symmetric under $\Delta(3n^2)$ and $\Delta(6n^2)$ groups and identifying their CP properties.

\ack

IdMV has received funding from Funda\c{c}\~{a}o para a Ci\^{e}ncia e a Tecnologia (FCT) through the
contract IF/00816/2015.
This project has received funding from the European Union's Seventh Framework Programme for research, technological development and demonstration under grant agreement no PIEF-GA-2012-327195 SIFT.

\section*{References}

\bibliography{ivo}

\end{document}